\documentclass[a4paper,fleqn]{cas-sc}

\usepackage{multirow}

\def\tsc#1{\csdef{#1}{\textsc{\lowercase{#1}}\xspace}}
\tsc{WGM}
\tsc{QE}
\tsc{EP}
\tsc{PMS}
\tsc{BEC}
\tsc{DE}

\usepackage{pmboxdraw}
\usepackage{lscape}
\usepackage[dvipsnames]{xcolor}
\usepackage{algorithm}
\usepackage{algorithmic}
\usepackage{caption}
\usepackage{changepage}
\usepackage{tabularx}
\usepackage{array}
\usepackage{natbib}

\usepackage{mathtools}

\setcitestyle{square}
\begin{document}
\let\WriteBookmarks\relax
\def\floatpagepagefraction{1}
\def\textpagefraction{.001}

\shorttitle{Quantum Cloud Computing: Trends and Challenges}

\shortauthors{Golec et al.}

\title [mode = title]{Quantum Cloud Computing: Trends and Challenges}

\author[1,2]{Muhammed Golec} [orcid=0000-0003-0146-9735]
\ead{m.golec@qmul.ac.uk}
\author[3]{Emir Sahin Hatay}
\author[4]{Mustafa Golec}
\author[5]{Murat Uyar}
\author[6]{Merve Golec}
\author[1]{Sukhpal Singh Gill}

\affiliation[1]{organization={Queen Mary University of London, United Kingdom}}
\affiliation[2]{organization={Abdullah Gul University, Kayseri, Turkey}}
\affiliation[3]{organization={University of Essex, United Kingdom}}
\affiliation[4]{organization={Kutahya Dumlupınar University, Kütahya, Turkey}}
\affiliation[5]{organization={Bursa Uludag University, Bursa, Turkey}}
\affiliation[6]{organization={Bursa Technical University, Bursa, Turkey}}

\cortext[cor2]{Correspondence to: School of Electronic Engineering and Computer Science, Queen Mary University of London, London, E1 4NS, UK.}

\begin{abstract} 
Quantum computing (QC) is a new paradigm that will revolutionize various areas of computing, especially cloud computing. QC, still in its infancy, is a costly technology capable of operating in highly isolated environments due to its rapid response to environmental factors. For this reason, it is still a challenging technology for researchers to reach. Integrating QC into an isolated remote server, like a cloud, and making it available to users can overcome these problems. Furthermore, experts predict that QC, with its ability to swiftly resolve complex and computationally intensive operations, will offer significant benefits in systems that process large amounts of data, like cloud computing. This article presents the vision and challenges for the quantum cloud computing (QCC) paradigm that will emerge with the integration of quantum and cloud computing. Next, we present the advantages of QC over classical computing applications. We analyze the effects of QC on cloud systems, such as cost, security, and scalability. Besides all of these advantages, we highlight research gaps in QCC, such as qubit stability and efficient resource allocation. This article identifies QCC's advantages and challenges for future research, highlighting research gaps.
\end{abstract}

\maketitle

\section{Introduction}

\textcolor{black}{The concept of quantum first entered the literature when it was discovered by atomic physicists in the early 1900s \cite{heim2020quantum}. The idea that a computer could be produced using quantum mechanics was first expressed by Richard Feynman in 1981 \cite{kandala2017hardware}. Since it was very difficult to maintain the stability of qubits operating based on quantum superposition and quantum entanglement, the first quantum computer prototypes began to emerge in the early 2000s only with the efforts of pioneering companies such as IBM \cite{gill2024quantum}. Quantum computers are still in the development phase, and it is predicted that they can perform a task in as little as 200 seconds, which would take 10,000 years for the world's best supercomputer \cite{sasaki2014practical}. For this reason, complex and long-lasting processes such as the behavior of molecules can be carried out very quickly with quantum-based computers. In addition, it is obvious that quantum computing (QC) will be one of the technologies that will shape the future as it begins to enter the military, civil, and commercial fields.}

\textcolor{black}{In addition to the advantages they offer, quantum computers require a highly insulated environment because they are sensitive to external factors such as heat, temperature, and noise \cite{bernstein2017post}. For this reason, quantum computers are still an expensive and difficult-to-stabilize technology for end users. The idea of researchers being able to access quantum computer resources by integrating quantum computers with cloud computing was put forward as an extremely bright idea. In this way, quantum computers can be placed in a highly isolated data center and provide service to the endpoints of the server. This model is called quantum cloud computing (QCC) and a platform service can be offered for researchers where they can apply quantum applications and algorithms \cite{lou2024quantum}.}

\subsection{Motivation and Contributions}

\textcolor{black}{With its superior computing capabilities compared to classical computers, QC has already created great excitement in academia and the private sector. Recent academic studies show that QC can be used to solve difficult problems in many fields such as health, chemistry, and physics \cite{gill2022quantum}. However, it is still a great challenge to access quantum computers as they require high cost and high isolation technology to produce. Research shows  \cite{gill2022quantum, gill2024quantum} that with the integration of quantum computers into cloud computing, the cost will decrease and the necessary isolation can be provided, so they can be offered to end users. Extensive research still needs to be conducted to identify trends and challenges that arise during the integration of these two paradigms.}

\textcolor{black}{This article is one of the first research articles on Quantum Cloud Computing, which emerged by integrating Quantum computing and Cloud computing. We can summarize the contributions of the article as follows:}

\begin{itemize}

\item  \textcolor{black}{We explain the concept of quantum cloud computing and aim to lay the groundwork for future research by explaining quantum cloud computing and its basic concepts,}

\item  \textcolor{black}{We explain quantum cloud computing trends and examine software tools, applications, and algorithms developed for the quantum cloud. In this way, we summarize the latest developments for the reader,}

\item  \textcolor{black}{We highlight challenges in quantum cloud computing and identify those that may arise during the integration of quantum and cloud computing paradigms. This way, we aim to provide guidance for future researchers.}

\end{itemize}

\subsection{Organization}

\textcolor{black}{The rest of this article is structured as follows. Section \ref{sec:sec2} introduces the reader to the background of QC and then examines the application areas of  QC. Section \ref{sec:sec3} explains the advantages of combining quantum and cloud computing and then concludes the paper by highlighting its applications, trends, and challenges.}

\section{Quantum Computing: Buzzword or Game Changer}\label{sec:sec2}

\begin{figure}[h]
    \centering
    \includegraphics[width= 0.6\linewidth]{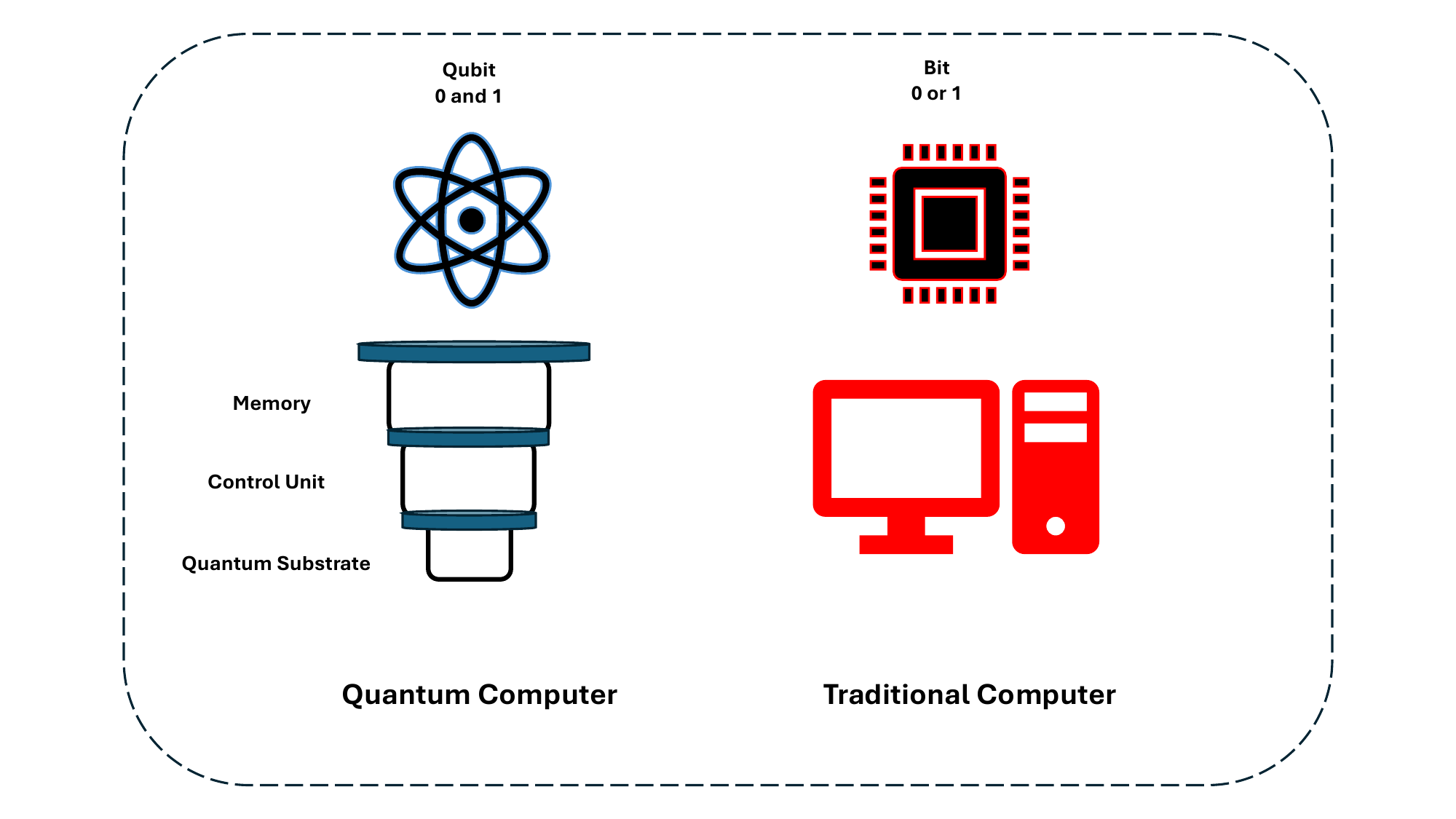}
    \caption{\textcolor{black}{Quantum Computing vs Traditional Computing.}}
    \label{fig:quantumvstraditional}
\end{figure}

\textcolor{black}{In computing systems, all operations such as data processing, communication, and storage are provided by bits consisting of 0 and 1 values \cite{gill2022ai}. Here, the one represented by 0 represents the low voltage in the electronic circuits, while the one represented by 1 represents the relatively high voltage in the electronic circuit. In today's computers, all operations are still performed with binary bit logic. QC, which has been frequently encountered in the academy and private sectors in recent years and is expected to shape the computing systems of the future, suggests that a bit can take values other than 0 and 1 \cite{brassard2003quantum}. Figure \ref{fig:quantumvstraditional} shows the working principle difference between QC and traditional computing.}
\textcolor{black}{QC works based on three basic principles found in quantum physics \cite{schwaller2021evidence}:}

\begin{itemize}

\item \textcolor{black}{\textit{Wave-particle Duality: }In quantum mechanics, particles of light, photons, are considered both waves and matter. In this way, a particle cannot be found only in one location but can be found everywhere.}

\item \textcolor{black}{\textit{Uncertainty Principle:} This principle, also called the Heisenberg Principle, means that the position and momentum of a photon cannot be measured at the same time.}

\item \textcolor{black}{\textit{Superposition:} In quantum mechanics, it means that the photon can be in more than one different position at the same time and it is impossible to determine its position. In this way, for computing operations such as communication and storage, a bit can take values other than 0 and 1, unlike in classical computing.}

\end{itemize}

\textcolor{black}{Quantum computing brings some advantages over classical computing \cite{biamonte2017quantum}:}

\begin{itemize}

\item \textcolor{black}{\textit{Speed:} QC has parallel processing capabilities such as performing multiple operations simultaneously. In this way, it works much faster than normal computers.}

\item \textcolor{black}{\textit{Security:} Since QC has the ability to break security algorithms such as RSA (Rivest–Shamir–Adleman) very quickly, it will lead to the development of the concept of Quantum Cryptography. This means much higher security.}

\item \textcolor{black}{\textit{Complex Problem Solving:} QC works more effectively in nonlinear and complex problems than classical computer systems, due to the features explained above due to the nature of qubits.}

\end{itemize}

\subsection{Algorithms and Software Tools}
\textcolor{black}{In this subsection, algorithms and software tools developed for QC will be examined. These algorithms and software tools can also be used for quantum cloud computing.}

\begin{figure}[h]
	\centering
	\includegraphics[width= 1\linewidth]{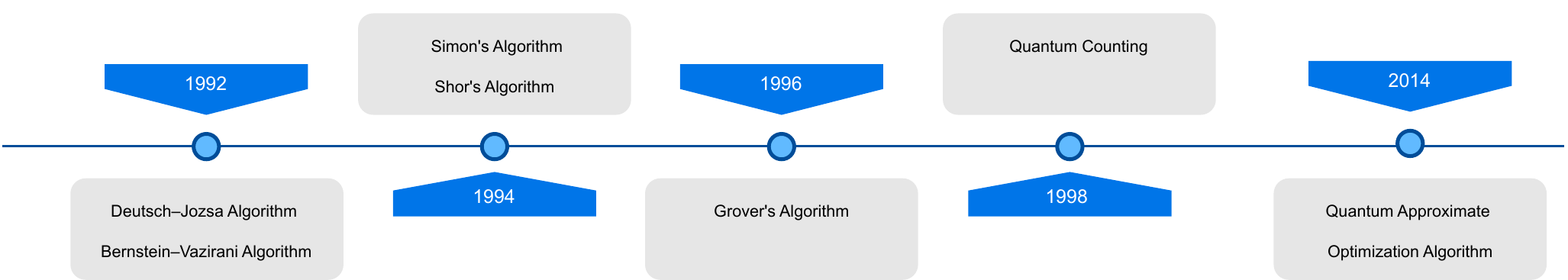}
	\caption{\textcolor{black}{The Timeline of Quantum Computing Algorithms.}}
	\label{fig:quantumalgos}
\end{figure}

\textbf{Algorithms}: \textcolor{black}{Figure \ref{fig:quantumalgos} shows popular quantum algorithms.} \textcolor{black}{ QC relies on the fundamental principles of quantum mechanics. The Deutsch-Jozsa algorithm is one of the first examples showing that a problem can be solved faster on quantum computers than on classical computers \cite{qiu2020revisiting}. It is a deterministic quantum algorithm discovered by David Deutsch and Richard Jozsa in 1992.  It emphasizes the advantage of employing negative amplitudes, which classical computers are incapable of accomplishing, in the Deutsch-Jozsa algorithm, a precursor to the development of far more major quantum algorithms. The Bernstein-Vazirani quantum algorithm, which is a limited version of the Deutsch-Jozsa algorithm developed by Ethan Bernstein and Umesh Vazirani in 1992, solves the hidden shift problem, which is important in error correction and cryptography \cite{nagata2017generalization}. Daniel Simon first published Simon's algorithm in 1994, which is superior to classical algorithms in terms of performance \cite{strubell2011introduction}. Simon's algorithm is employed to solve the black box function that fulfills a given collection of values and solves it at a much more exponential speed than any classical algorithm by using a quantum computer. Simon's algorithm served as the inspiration for Shor's algorithm, which is a quantum algorithm developed by Peter Shor in 1994 for the purpose of finding the prime factors of an integer. After these algorithms which were based on the Fourier transform, Grover's algorithm is a quantum algorithm that offers a better improvement than classical algorithms for unstructured search problems developed based on amplitude amplification, which enables quantum computers capable of quickly solving challenges that may be difficult to solve with classical computers and quantum counting based on it was developed respectively \cite{cerezo2021variational}. Finally, quantum approximate optimization, a hybrid quantum-classical algorithm that focuses on solving graph theoretic problems and is predicted to obtain better solutions than classical algorithms, has recently been introduced \cite{zhou2020quantum}.}

\begin{figure}[h]
    \centering
    \includegraphics[width= 0.6\linewidth]{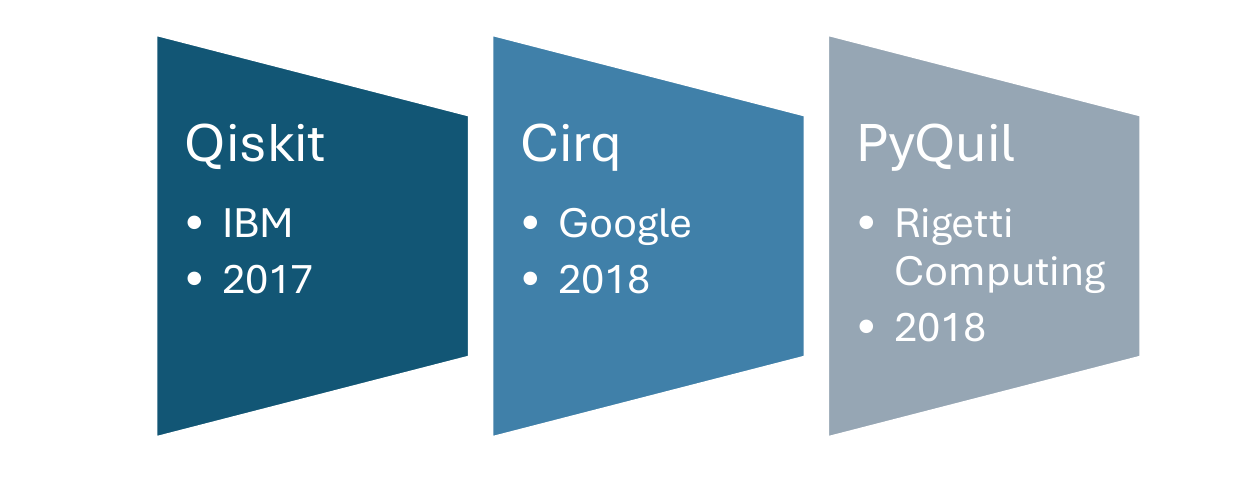}
    \caption{\textcolor{black}{Software Tools for Quantum Cloud Computing .}}
    \label{fig:quantumsoftware}
\end{figure}

\textbf{Software Tools}: \textcolor{black}{Figure \ref{fig:quantumsoftware} shows popular quantum software tools, which can be utilized for quantum cloud computing. \textcolor{black}{ Quantum software is an arising and less developed topic than quantum technology. The software that manages the quantum hardware is expected to be able to utilize complex quantum techniques and deliver high-level performance. In these days, there are different platforms for quantum computer programming depending on hardware solutions. One of these is Qiskit (Quantum Information Science Kit), a software development kit developed by IBM in 2017 for the operation of quantum computers at the algorithm and circuit level \cite{wille2019ibm}. Another is Cirq, an open-source framework developed by Google for quantum computers \cite{hancockcirq}. PyQuil, which allows running programs on real quantum computers using the Quantum Cloud Service, was developed by Rigetti Computing \cite{hibat2024framework}. These frameworks are frequently made available by quantum developers under open-source and with API (Application Programming Interface) in Python.}}

\subsection{Research Gaps and New Trends} \label{subsec:trens}

\begin{figure}[h]
    \centering
    \includegraphics[width=0.5\linewidth]{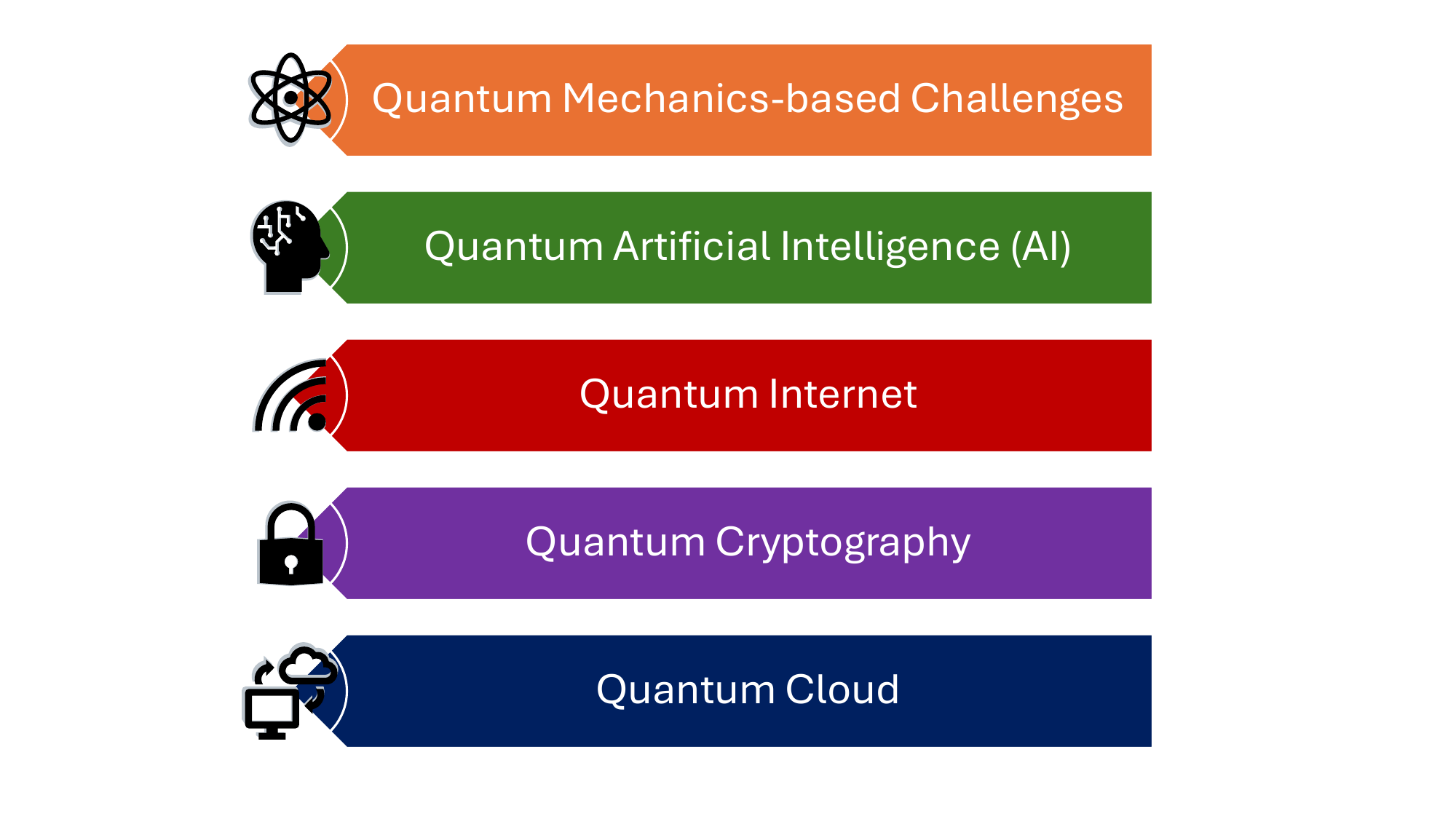}
    \caption{\textcolor{black}{Research Gaps and New Trends.}}
    \label{fig:quantumtrends}
\end{figure}

\textcolor{black}{In this subsection, we examine the research gaps and new trends related to QC. Figure \ref{fig:quantumtrends} highlights these research gaps regarding QC for researchers.}

\begin{itemize}
\item \textcolor{black}{\textbf{Quantum Mechanics-based Challenges} \cite{imre2014quantum}: Challenges arising from quantum mechanics have still not been overcome in quantum computers. These difficulties can generally be examined under two subheadings: i) Qubits lose data due to the situation called decoherence in quantum mechanics. This causes the coherence time to be short and is undesirable in QC applications. ii) Another challenge is the quantum error correction. Compared to classical computers, error correction is a difficult task in quantum computers due to reasons such as the more complex errors and the difficulty of copying the quantum state.}
\item \textcolor{black}{\textbf{Quantum Artificial Intelligence (QAI) }\cite{dunjko2018machine}: By combining Machine Learning (ML) \& Deep Learning (DL) methods with quantum processing power, difficult and complex problems can be solved more effectively. However, in quantum computers that still contain a high number of qubits (millions), problems such as internal noise can reduce the performance of the models.}
\item \textcolor{black}{\textbf{Quantum Internet} \cite{wehner2018quantum}: Unlike the traditional Internet, communication is achieved using quantum principles such as superposition and entanglement. In this way, key distribution difficulties in cryptography are prevented and communication security is increased. In addition to this advantage, difficulties such as qubit fragility, transmission, and long-distance data transmission (entanglement dispersion) arising from quantum-based data need to be solved.} 
\item \textcolor{black}{\textbf{Quantum Cryptography} \cite{kumar2022securing}: It is thought that modern cryptography methods will be endangered with the proliferation of quantum computers with high processing capabilities. For this reason, new measures based on quantum principles are needed to ensure secure key distribution and encryption. Based on quantum principles, Quantum Cryptography can prevent a hacker from making a copy and monitoring the key in the encryption algorithm. However, there is still a need for new Quantum Cryptography protocols that minimize the impact of qubits from environmental factors and prevent data loss due to decoherence during transmission.}
\item \textcolor{black}{\textbf{Quantum Cloud} \cite{lou2024quantum}: Since quantum computers are expensive and require a high level of isolation due to qubit instability, it is a logical solution to make them available to users through a central server. However, intensive studies are required on this concept that combines QC and cloud computing. Providing quantum processing power to researchers through cloud platforms will provide great advantages for developing quantum algorithms, quantum simulations, and other quantum-based applications. However, security, scalability (in line with increasing demand), infrastructure, and software studies still need to be carried out for quantum and cloud integration.}
\end{itemize}

\section{Quantum Cloud Computing}\label{sec:sec3}

\begin{figure}[h]
    \centering
    \includegraphics[width=0.8\linewidth]{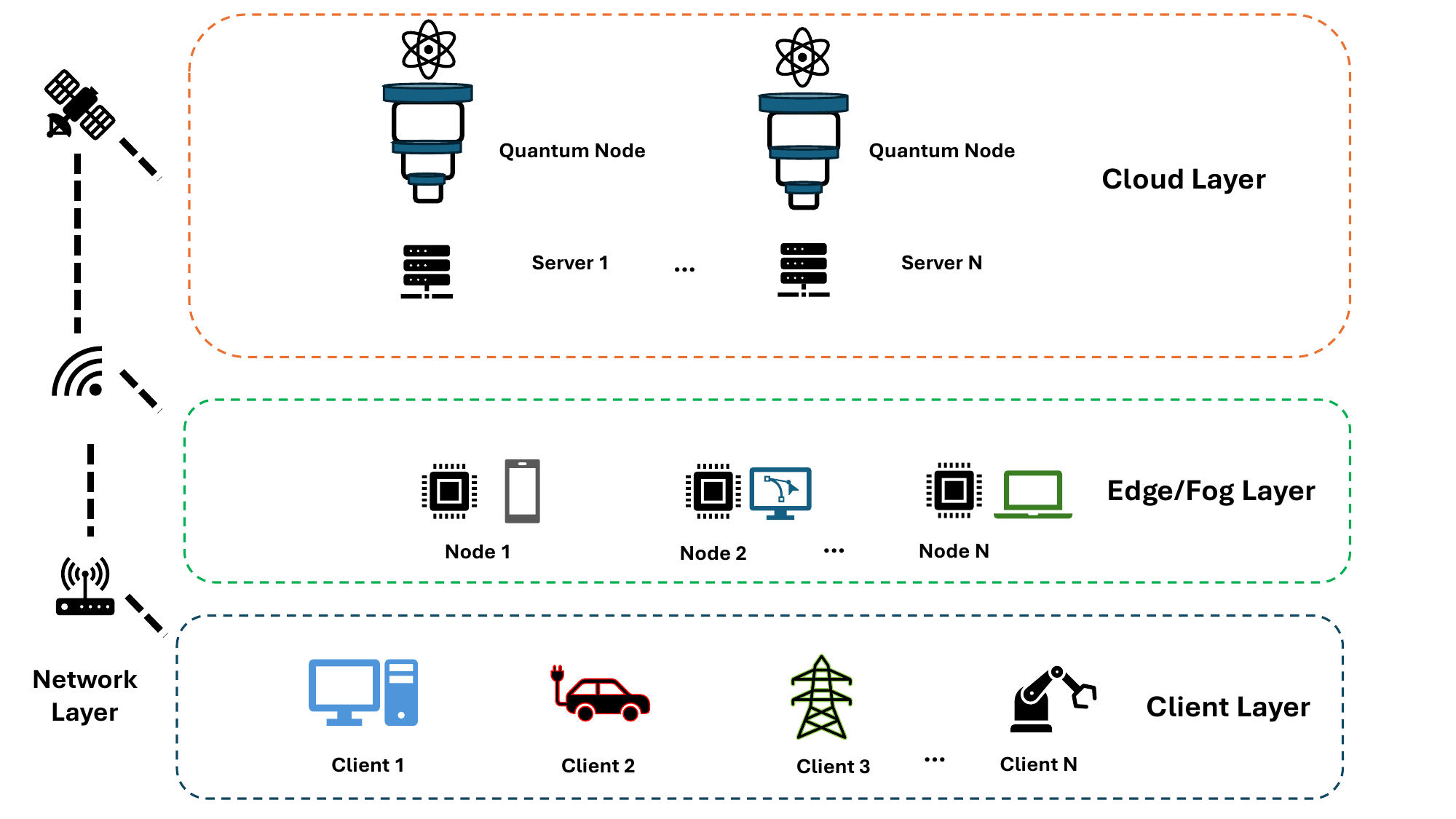}
    \caption{\textcolor{black}{General Architecture of Quantum Cloud Computing.}}
    \label{fig:quantumcloudmain}
\end{figure}

\textcolor{black}{This section examines the basic concepts, trends, applications, and challenges of quantum cloud computing (QCC). The concept of QCC is a new computing paradigm that aims to facilitate end-user access to QC by using cloud computing platforms. Thanks to QCC, users will have easy access to quantum computers, which are costly and require high stabilization. Figure \ref{fig:quantumcloudmain} shows a general architecture of QCC. The Quantum computer can be placed on a cloud-based platform and serve the end user via an Application Programming Interface (API). Likewise, quantum processing power can be distributed to nodes such as edge and fog, which reduces latency and bandwidth traffic. Cloud platforms provide the environment parameters (network infrastructure, storage, operating environment, etc.) required for the quantum computer and undertake the task of transmitting the calculation results on quantum computers to the farthest point of the network. The Network Layer shown in the figure shows the computing technologies currently used in server-client communication. With the development of quantum internet technologies explained in Section \ref{subsec:trens}, major improvements such as throughput and latency are expected to be achieved in data communication.}

\subsection{New Research Paradigm}
\textcolor{black}{Work on QCC, which is believed to be one of the most popular uses of QC in the future, continues at a great pace. Leading cloud companies such as Amazon (Braket), Microsoft (Quantum Development Kit), and IBM (Quantum Experience) have now started to offer quantum services to the public, albeit with limited processing power \cite{mehta2023survey}. These initiatives, which are still in their infancy, provide clues that the accessibility of QCC will become even easier in the future  \cite{gill2024quantumcomputing, singh2022quantum, gill2024modern}. The advantages of QCC can be listed as follows:}

\begin{itemize}
     
\item \textcolor{black}{Cloud-based QC platforms enable users to rapidly expand their computing capacity according to their needs. This feature enables the allocation of suitable resources for QC issues of varying magnitudes.}

\item \textcolor{black}{Cloud-based QC training programs can be used to introduce quantum circuits and better understand the advantages of QC.}

\item \textcolor{black}{Cloud-based QC can develop and test quantum algorithms before running them on actual quantum computers. In this way, the necessary skills and expertise are not needed to access quantum computers and Cloud-based QC reduces the costs of cloud platforms.}

\item \textcolor{black}{It ensures that everyone has access to the same systems and makes it easier for individuals with Internet connections from all over the world to collaborate.}

\item \textcolor{black}{QCC platforms can enable users to benefit from the expertise of quantum experts provided by the service platform. Thus, users or organizations new to QC can develop quantum calculations and run them on real quantum computers.}

\item \textcolor{black}{It reduces the need for physical security measures that would be necessary if pricey technology were retained on-site.}

\end{itemize}

\subsection{Applications and Future Trends}

\begin{figure}[h]
    \centering
    \includegraphics[width=0.76\linewidth]{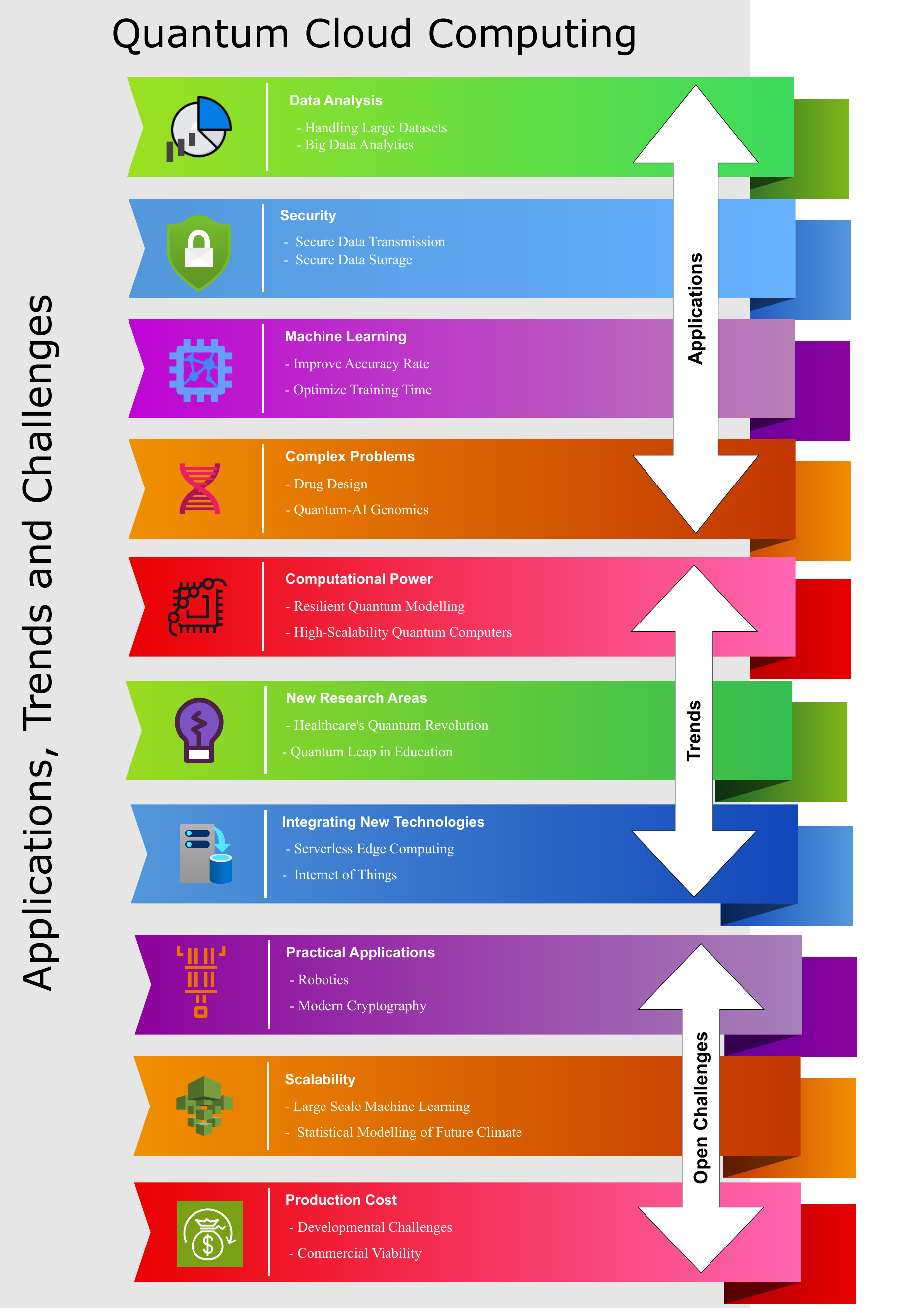}
    \caption{\textcolor{black}{Applications, Future Trends and Challenges of Quantum Cloud Computing. }}
    \label{fig:quantuchallenge}
\end{figure}

\textcolor{black}{In this subsection, possible applications and future trends of QCC are examined. Figure \ref{fig:quantuchallenge} shows the applications, trends, and challenges of QCC. By taking advantage of a quantum computer inside, QCC can solve tasks where the processing power of traditional computers is insufficient or processes that take a long time with a traditional computer, in a very short time  \cite{gill2024quantumcomputing, singh2022quantum, gill2024modern}. Therefore, its future application areas are very wide. }

\textbf{Applications:} Some of the highlights of these areas are: 

\begin{itemize}
  \item  \textcolor{black}{\textit{Data Analysis:} It can be used in applications containing large data sets,   }
  
  \item  \textcolor{black}{\textit{Security:} It can be used to transmit and store data securely in the period when quantum computers become widespread,  } 
  
  \item  \textcolor{black}{\textit{Machine Learning:} It can shorten the training time of ML and DL models and can seriously increase the accuracy rates of the models, }
  
  \item  \textcolor{black}{\textit{Complex Problems:} It can be used for innovative solutions in complex and difficult technologies such as drug design and gene technologies.}

\end{itemize}

\textcolor{black}{\textbf{Future Trends:} Little is still known about QCC application areas, leaving major research gaps for many researchers. Possible trends regarding QCC are as follows: }

\begin{itemize}

\item \textcolor{black}{Increasing the power of quantum computers for solving more complex problems and AI models involving large datasets, }

\item \textcolor{black}{Expanding QCC to broader areas such as health and education by reducing costs, }

\item \textcolor{black}{Integrating quantum computing with the cloud raises questions about infrastructure development and security standards.}

\end{itemize}

\subsection{Open Challenges}

\textcolor{black}{Although QCC is a promising paradigm, it still has some challenges to be solved before it can be adopted \cite{gill2024quantumcomputing, singh2022quantum, gill2024modern}. The most important of these challenges are as follows:}
\begin{itemize}

    \item  \textcolor{black}{Quantum computers are still under development and cannot be used for practical applications. }
    
    \item \textcolor{black}{In cloud computing, resources can be scaled to meet demand fluctuations. Quantum computers are difficult to manage on a large scale and may not meet the scalability expected in cloud services.} 
    
    \item \textcolor{black}{Since producing quantum computers still requires high costs, QCC systems will also cause higher costs than traditional cloud services.}
    
\end{itemize}

\section{Conclusions}

\textcolor{black}{Quantum Computing (QC) is a new paradigm that is still in its infancy but has the potential to revolutionize many scientific and technological fields in the future. Due to quantum mechanics, quantum computers operate in well-insulated environments, making them susceptible to environmental factors like heat, temperature, and noise. As a result, it is still a difficult technology for quantum researchers to reach. By integrating QC into cloud platforms, providing service to the network's edge from a well-isolated server can reduce costs. This brilliant idea may be possible with quantum cloud computing (QCC), which emerges with the integration of QC and cloud computing. This article aims to provide guidance for researchers by highlighting the QCC paradigm, future application areas, advantages, challenges, and research gaps.}

\section*{\uppercase{Acknowledgements}}

Muhammed Golec would express his thanks to the Ministry of Education of the Turkish Republic, for their support and funding.


\bibliographystyle{apalike}

\bibliography{cas-refs}

\end{document}